\documentclass[prb,twocolumn,showpacs,superscriptaddress,preprintnumbers,amsmath,amssymb]{revtex4}

\usepackage{graphicx}%
\usepackage{dcolumn}%

\begin{document}

\title{Measuring the gap in ARPES experiments}

\author{A. A. Kordyuk}
\affiliation{Institute for Solid State Research, IFW Dresden, P.O. Box 270016, D-01171 Dresden, Germany}
\affiliation{Institute of Metal Physics of National Academy of Sciences of Ukraine, 03142 Kyiv, Ukraine}

\author{S. V. Borisenko}
\author{M. Knupfer}
\author{J. Fink}
\affiliation{Institute for Solid State Research, IFW Dresden, P.O. Box 270016, D-01171 Dresden, Germany}

\date{August 21, 2002}%

\begin{abstract}
Angle-resolved photoemission spectroscopy (ARPES) is considered as the only experimental tool from which the momentum distribution of both the superconducting and pseudo-gap can be quantitatively derived. The binding energy of the leading edge of the photoemission spectrum, usually called the leading edge gap (LEG), is the model-independent quantity which can be measured in the modern ARPES experiments with the very high accuracy---better than 1 meV. This, however, may be useless as long as the relation between the LEG and the real gap is unknown. We present a systematic study of the LEG as a function of a number of physical and experimental parameters. The \textit{absolute} gap values which have been derived from the numerical simulation prove, for example that the nodal direction in the underdoped Bi-2212 in superconducting state is really the node---the gap is \textit{zero}. The other consequences of the simulations are discussed.
\end{abstract}


\maketitle

\section{Introduction}

The question of symmetry of the superconducting gap is crucial to understand the nature of superconductivity in cuprates.\cite{TsueiRMP00} Although not being phase sensitive, the angle-resolved photoemission spectroscopy (ARPES) has been considered as the only experimental tool from which both the superconducting and pseudo-gap anisotropy could be quantitatively derived.\cite{ShenPRL93, DingPRL95, MarshallPRL96, DingPRL97, MesotPRL99} The binding energy of the leading edge of the photoemission spectrum, usually called the \textquotedblleft leading edge gap" (LEG),\cite{MarshallPRL96, DingPRL97} is the model-independent quantity which can be measured in the modern ARPES experiments with the highest, up to 1 meV accuracy,\cite{BorisXXX02} and is, actually, the best quantity from which we can judge about the real gap values---in absence of a widely accepted model for the gap formation in these compounds, there is no direct way to extract the real gap value from the ARPES spectra.

The LEG is determined as a lowest binding energy at which the energy distribution curve (EDC) reaches half of its maximum (here and thereafter we keep the same notation \textquotedblleft LEG" for either gapped or non-gapped spectra). Therefore it is understandable, and usually admitted, that the LEG should depend on any parameters that determine the EDC line shape and, unless the relation between the LEG and the real gap is known, can be considered only as a \textit{qualitative} representation of the real gap. Moreover, the mentioned parameters (and EDC line shape) are substantially changing during any experiment, e.g.~with temperature, along the Fermi surface (FS), etc. So, analysing the experimental data it is very important to distinguish between the artificial variations of the LEG and changes caused by the real gap in the electronic density of states. Despite the big importance of this question, there is no systematic study of the LEG available today. Hence, even the qualitative relation between the LEG and the real gap can be doubtful. 

In this paper, by means of a numerical simulation we present a systematic investigation of the LEG as a function of a number of physical and experimental parameters which govern the ARPES spectra of the Bi-2212 cuprates. 

Starting with the non-gapped case, we demonstrate that the dependencies of the LEG on the temperature and slight deviations in the momentum space (even within the momentum resolution window) from the FS ($\textbf{k}_F$) result in strong variations of the LEG values which are comparable with the values of the superconducting and pseudo-gaps. We also show that at the low temperatures, choosing an appropriate criterion for the $k_F$ determination---the \textquotedblleft maximum intensity" or \textquotedblleft minimum gap locus"---the absolute value of the LEG is not very sensitive to the \textquotedblleft physical" parameters of the model spectral function and therefore can be easily obtained. As an example we consider the experimental data of the LEG for an underdoped Bi-2212 in the superconducting state,\cite{BorisXXX02} and prove that the nodal direction there is really the node---the gap is \textit{zero} within the experimental accuracy. Finally, switching to the gapped case, we show that the absence of a \textquotedblleft cusp" of the LEG vs FS angle dependencies (so called \textquotedblleft U-shape"),\cite{BorisXXX02} cannot be explained neither by the interplay between the temperature and Fermi-function nor by other \textquotedblleft artificial" parameters which may vary over the FS: momentum resolution, self-energy and band-structure. As an intermediate result, we present the tight-binding parameters for the considered compounds.

\section{Model description}

We start with a model spectral function
\begin{eqnarray}\label{A0}
A(\omega,\epsilon,T) \propto \frac{|\Sigma''(\omega,T)|}{(\omega - \epsilon - \Sigma'(\omega,T))^2 + \Sigma''(\omega,T)^2},
\end{eqnarray}
which is an essential part of the photocurrent measured in the experiments 
\begin{eqnarray}\label{photocurrent}
I(\textbf{k},\omega,T,h\nu) \propto [(M(\textbf{k},h\nu) A(\textbf{k},\omega,T) f(\omega,T)]\nonumber\\ \otimes~R_{\textbf{k}} \otimes~R_{\omega} + B(\omega,T).
\end{eqnarray}
Here $M(\textbf{k},h\nu)$ represents the dependence of the squared one-electron matrix element on in-plane electron momentum \textbf{k} and excitation energy $h\nu$, $f(\omega,T) = 1 / [\exp(\omega/T)+1]$ is the Fermi function. $R_{\textbf{k}}$ and $R_{\omega}$ represent the functions of the momentum and energy resolutions respectively. We include the energy resolution as a convolution with a Gaussian of the $R_\omega$ full width at half maximum (FWHM) and the momentum resolution through a simple summation over a part of the detector area,
\begin{eqnarray}\label{Resk}
A \otimes R_{\textbf{k}} = \sum_{\delta\textbf{k}} A,
\end{eqnarray} 
where $\delta\textbf{k}$, in general case, is a 2D area with the $R_{k\parallel} \times R_{k\perp}$ dimensions (momentum resolutions parallel and perpendicular to the entrance slit of an analyser respectively) whose orientation in $\textbf{k}$-space depends on $\textbf{k}$ and the experimental geometry but, in the case of the big FS of superconducting cuprates, one can consider $\delta\textbf{k}$ as an one-dimensional cut $\pm R_{k}(\textbf{k})/2$ perpendicular to the FS.

For the momentum independent background we assume an empirical relation: $B(\omega, T) \propto (1 + b\omega^2) f(\omega - \Delta_b, T + T_{b})$ with $b$ = 1 eV$^{-2}$, $\Delta_b$ = 5 meV and $T_{b}$ = 90 K. The parameters $\Delta_b$ and $T_{b}$ depend on doping and $\Delta_b$ also depends on temperature \cite{KordPRL} but we neglect this here because, as it is shown below, the typical signal-to-background ratios are large and the influence of the background on the LEG value is rather weak.

\begin{figure*}
\includegraphics[width=18.4cm]{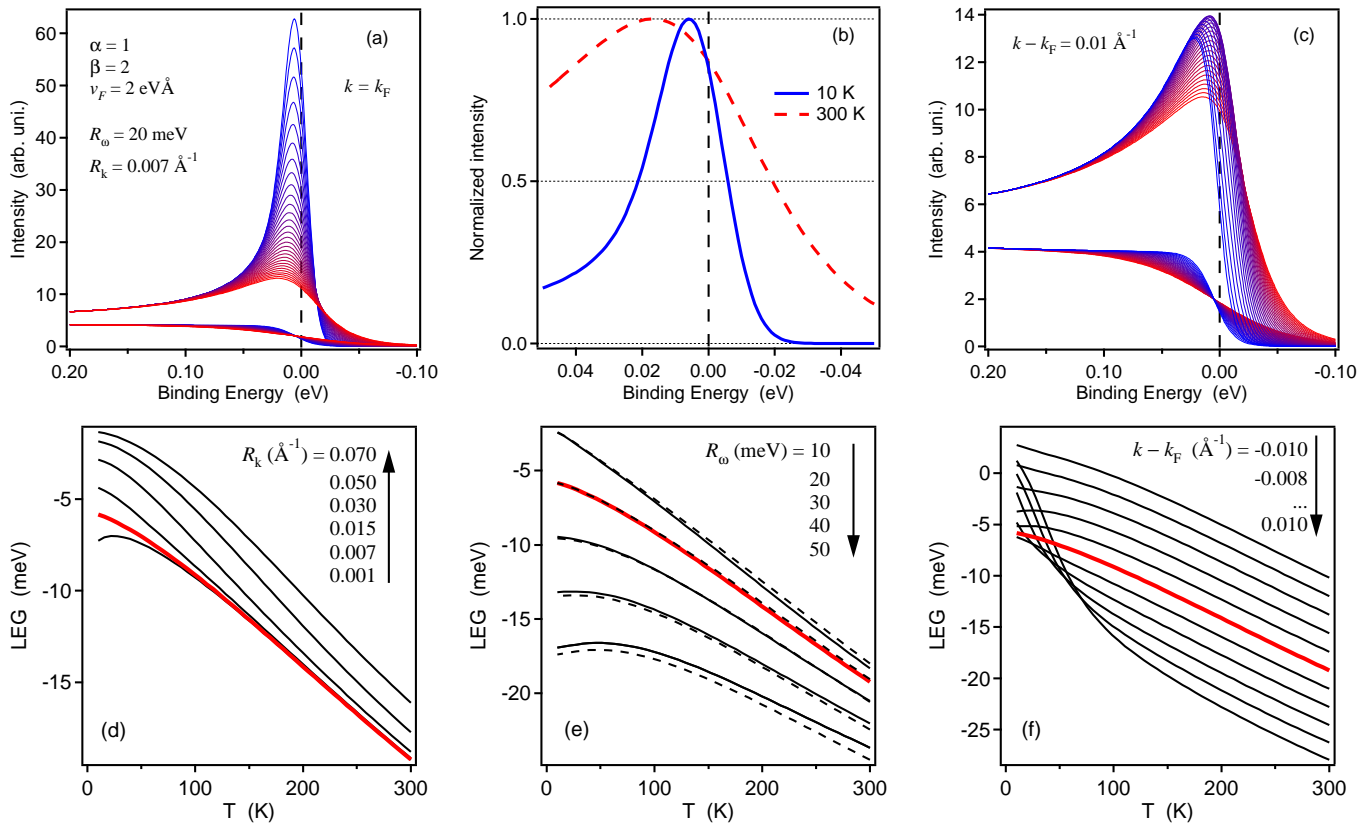}%
\caption{\label{Tdep} Temperature dependence of the model photoemission spectra with no gap included (\textit{a--c}: temperatures from 10 to 300 K shown in color from blue to red with 10 K step) and the leading edge gap (LEG) values derived therefrom for different momentum (\textit{d}) and energy (\textit{e}) resolutions and positions in $k$ (\textit{f}). The lower curves on panels \textit{a} and \textit{c} represent the temperature dependent background. The red bold curves on panels \textit{d--f} show the LEG for typical experimantal parameters $R_{\omega}$ = 20 meV, $R_k$ = 0.07~\AA$^{-1}$, $k = k_F$.}
\end{figure*}

For the imaginary part of the self energy we also use an empirical formula \cite{Self-energy} $\Sigma''(\omega,T) = \sqrt{(\alpha_0 \omega)^2 + (\beta_0 T)^2}$ which fits best the experimental data (see Refs.~\onlinecite{BorisPRB01,KordPRL}). In the vicinity of the Fermi-level ($\pm$ 50 meV) the real part of the self-energy can be well approximated by its linear term $\Sigma'(\omega,T) \approx -\lambda(T) \omega$ ($\lambda > 0$) and the spectral function (\ref{A0}) can be rewritten in a renormalized form
\begin{eqnarray}\label{A1}
A(\omega,\textbf{k},T) \propto \frac{1}{1+\lambda} \frac{|\Sigma''(\omega,T)|}{(\omega - \varepsilon(\textbf{k}))^2 + \Sigma''(\omega,T)^2},
\end{eqnarray}
with the renormalized imaginary part of the self-energy $\Sigma'' = \sqrt{(\alpha \omega)^2 + (\beta T)^2}$, $\alpha = \alpha_0/(1+\lambda)$, $\beta = \beta_0/(1+\lambda)$, and the renormalized dispersion $\varepsilon = \epsilon /(1+\lambda)$. The last one, in a small vicinity of the Fermi energy $E_F$ on the path perpendicular to FS, can be written using the renormalized Fermi velocity, $\varepsilon = v_F k$, which in the same way relates to the bare one: $v_F = u_F/(1+\lambda)$.

At last, when we take into account the bilayer splitting we include it as a simple superposition of the photocurrent from bonding (\textquotedblleft\textit{b}\textquotedblright) and antibonding (\textquotedblleft\textit{a}\textquotedblright) bands:
\begin{eqnarray}\label{splitting}
I \propto [M_a A(\varepsilon_a)+ M_b A(\varepsilon_b)] f \otimes R_{\textbf{k}} \otimes R_{\omega} + B.
\end{eqnarray}

\section{Gapless case}

First, we examine the LEG behaviour assuming that there is no real gap at all. It is natural to expect that the leading edge position of an EDC should depend on every physical parameter which forms the spectral function and the background as well as on energy and momentum resolutions. Among others, the temperature seems to be the most crucial here (see Ref.~\onlinecite{BorisPRB01}).   

\subsection{Temperature dependence}

Fig.~\ref{Tdep} shows the temperature dependence of the model photoemission spectra (\textit{a--c}) defined by Eqs.~\ref{photocurrent} and \ref{A1} and the LEG values (\textit{d--f}) derived therefrom. Here we use the following parameters: $\alpha$ = 1, $\beta$ = 2, $v_F$ = 2~eV\AA, $R_{\omega}$ = 20 meV, and $R_k$ = 0.07~\AA$^{-1}$ that corresponds to 0.2$^{\circ}$ angular resolution for the nodal point at 21 eV excitation energy. Panel (\textit{a}) represents the photocurrent (\ref{photocurrent}) at $k = k_F$ ($k_F$-EDCs) at different temperatures from 10 to 300 K (shown in color from blue to red with 10 K step). Panel (\textit{b}) shows two extreme spectra but normalized to their maximum in order to illustrate the motion of the leading edge with temperature. The LEG($T$) dependencies are shown on panel (\textit{d}) and (\textit{e}) for different momentum and energy resolutions.

The signal-to-background ratio (which can be estimated as a ratio of peak-to-tail photocurrent values) for the EDCs presented on panel (\textit{a}) is chosen to be similar to typical photoemission data, and panel (\textit{e}) also shows that the influence of the background in this case is negligible---the LEG($T$) for the spectra without the background are shown as dashed lines.

It can be seen that, although the LEG strongly depends on temperature and moves about 15 meV going from 10 to 300 K, which should be taken into account when the temperature dependence of the real gap is studied, this dependence, for reasonable energy and momentum resolutions ($R_{\omega}$ = 20 meV, $R_k$ = 0.07~\AA$^{-1}$: shown as a bold red curve in Fig.~\ref{Tdep}, \textit{d--f}), is quite monotonic (i.e. has no maxima on both itself and its derivative) and cannot imitate a rapid gap opening at a certain temperature.

The situation is different if EDCs are taken from a $k$-point which does not exactly coincide with $k_F$. Panels (\textit{c}) and (\textit{f}) of Fig.~\ref{Tdep} illustrate this case. Panel (\textit{c}) shows EDCs for $k = k_F$ + 0.01~\AA$^{-1}$ for the same temperature range (10--300 K) and panel (\textit{f}) shows the LEG($T$) dependencies for different $k-k_F$ (from -0.01 to 0.01~\AA$^{-1}$). One can see that the unoccupied part of the Brillouin zone ($k > k_F$) is the most dangerous in this sense. If one steps away from $k_F$ in the unoccupied direction in 0.01~\AA$^{-1}$ (about 0.3$^{\circ}$ for the nodal point at 21 eV excitation energy), the LEG changes about 30 meV going from 10 to 300 K but, what is most important, a kink appears at about 80 K on the LEG vs $T$ dependence. This kink \textit{can be} easily misinterpreted as a gap opening, so we may conclude that the uncertainty in $k_F$ determination (see discussion in Ref.~\onlinecite{BorisPRB01} and below) is the most crucial parameter for the correct LEG evaluation.

Regarding Fig.~\ref{Tdep} (\textit{a}) we also would like to note that due to the temperature dependent self-energy the area under each EDC is not constant with temperature, as one could expect according to the corresponding sum rule (e.g., see Ref.~\onlinecite{RanderiaPRL95}). 

\begin{figure*}
\includegraphics[width=6.5cm]{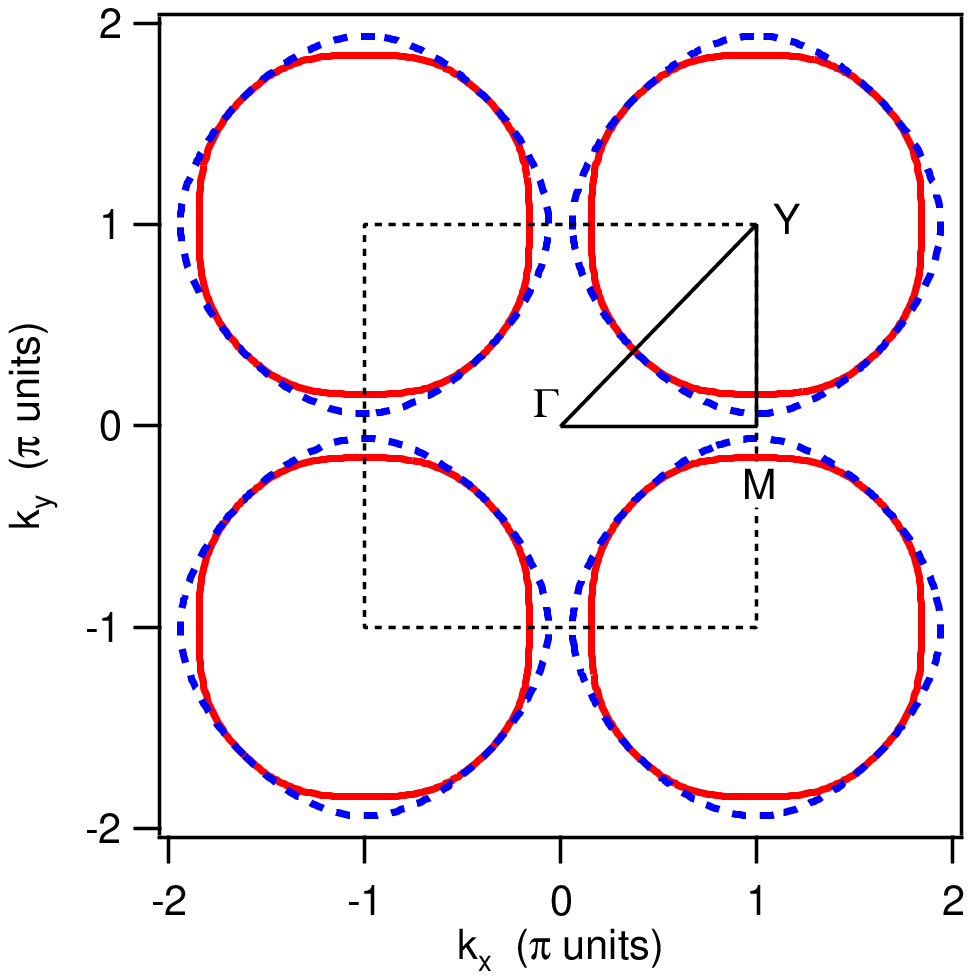}%
\includegraphics[width=10.9cm]{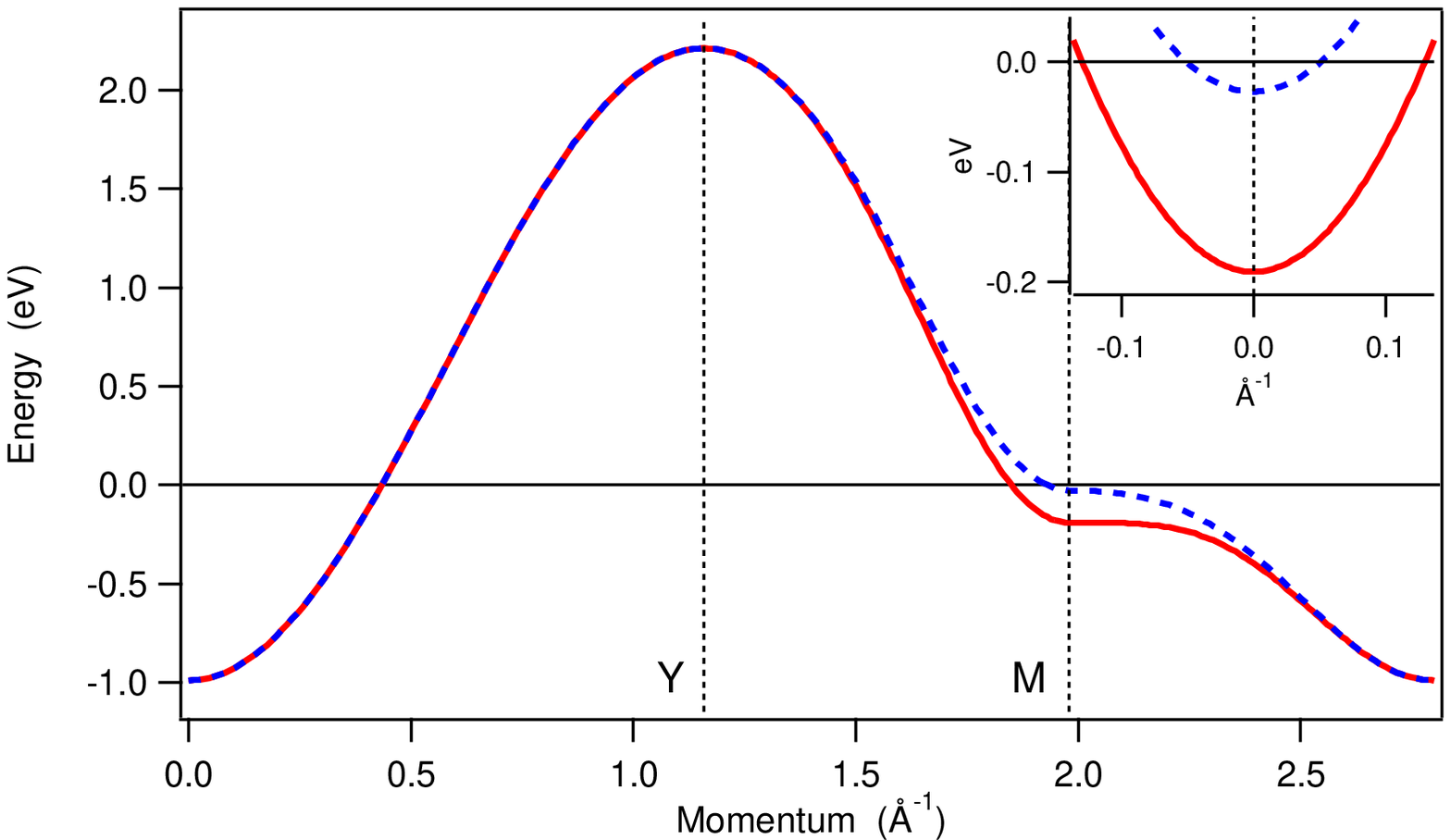}%
\caption{\label{TBF} The result of the tight-binding fit to the bonding (red solid lines) and antibonding (blue dashed lines) CuO-bilayer bands of the overdoped (OD 69K) Bi-2212: left panel shows the corresponding Fermi surfaces (the boundary of the first Brillouin zone marked by dotted square), right panel shows the 'bare' dispersion along the $\Gamma$-Y-M-$\Gamma$ path (shown in the left panel as a triangle), and inset zooms in the ($\pi$,0) region, from ($\pi$,$\pi/6$) to ($\pi$,$-\pi/6$).}
\end{figure*}

\subsection{Resolution}

The dependence of the LEG on resolutions can be estimated from Fig.~\ref{Tdep} (\textit{d}) and (\textit{e}). The dependence on energy resolution (Fig.~\ref{Tdep} (\textit{e})) is stronger but it is not a problem because we believe that normally it remains constant during the experiment and the absolute value can be easily taken into account. The momentum resolution \textit{can} change during the experiment. For example, for the experimental geometry described in Ref.~\onlinecite{BorisPRB01}, the actual momentum resolution changes going from the $\Gamma$--$(\pi,\pi)$ to $(\pi,\pi)$--$(\pi,0)$ FS crossings: it is defined by the angular resolution along the entrance slit of the analyser for the former and by the angular resolution perpendicular to the slit for the latter. As a result of this, the related momentum resolution can change by a factor of two but, as one can see from Fig.~\ref{Tdep} (\textit{d}), this will move the leading edge about 1 meV only (even for the "worst" case of 10~K).

\subsection{Band structure}

In contrast to the temperature, the other parameters that enter the $k_F$-EDCs are expected to depend on the in-plane momentum of the electrons. One can divide these parameters into \textquotedblleft structural\textquotedblright, which come from the bare electronic structure, and \textquotedblleft physical\textquotedblright, which come from interactions and reveal themselves as a self-energy.

The structural parameters of interest here are the bare Fermi velocity and the bilayer splitting. Fig.~\ref{TBF} represents the result of the tight-binding fit of the experimental data published in Refs.~\onlinecite{BorisXXX02} and \onlinecite{KordPRB02}. The precisely determined FS of Bi-2212 (at least its bonding sheet, see Ref.~\onlinecite{KordPRB02}) at room temperature (300 K) can be well described within the four-band model introduced in Ref.~\onlinecite{AndersenJPCS95}. Within this model, the simplified bare dispersion
\begin{eqnarray}\label{TBDispersion}
\epsilon_{a,b} & = & \Delta \epsilon - 2t (\cos k_x + \cos k_y) \nonumber\\
& + & 4t' \cos k_x \cos k_y - 2t'' (\cos 2k_x + \cos 2k_y) \nonumber\\
& \pm & t_{\perp} (\cos k_x - \cos k_y)^2 /4,
\end{eqnarray}
where second and third nearest-neighbor intra-plane hopping integrals ($t'$ and $t''$) are provided by the "Cu $s$" orbital and interlayer hopping is described by the $t_{\perp}$ (the bilayer splitting, the energy distance between bonding and antibonding CuO-bilayer bands, is $2 t_{\perp}$).\cite{AndersenJPCS95, LiechtensteinPRB96}

To fit the data we use the following procedure. First, the relative parameters (relative to $t$) are determined from the room temperature FS maps of Ref.~\onlinecite{KordPRB02}, then the bilayer splitting is taken into account [the value $t_{\perp}/t$ is determined from the $(\pi,\pi)-(\pi,0)$ FS crossing]. For example, for the overdoped sample (OD, $T_c$ = 69 K), using the assumption that $t'' \approx t'/2$,\cite{PavariniPRL01} these parameters are: $t'/t$ = 0.23, $t''/t$ = 0.11, $t_{\perp}/t$ = 0.21, $\Delta \epsilon / t$ = 1.08. From here one can estimate the variation of the bare Fermi velocity $u_F$ around the FS as a ratio of its value in the antinodal 'A' point (FS crossing along $(\pi,\pi)-(\pi,0)$ direction) to the nodal ($(0,0)-(\pi,\pi)$ crossing) one. This ratio, $u_F$(A)/$u_F$(N), is 0.71 for the bonding band and 0.25 for the antibondig band (for OD 69 K). We stress that these values are determined by the FS shape, as presented on the left panel of Fig.~\ref{TBF}, and do not depend on the energy scale.

In order to estimate the \textquotedblleft scale\textquotedblright~$t$ one needs to know some absolute value of the bare band. We can estimate the 'bare' Fermi velocity as 4.0 eV\AA, subtracting the real part of the self-energy, which has been determined using the Kramers-Kr\"onig transform of its imaginary part, from the renormalized experimental dispersion. The tight-binding parameters which have been calculated for two samples with this value of $u_F$, assuming the constancy of the bilayer splitting,\cite{ChuangXXX01} are presented in Table \ref{tab}. The right panel of Fig.~\ref{TBF} shows the obtained bare dispersion along the $\Gamma$-Y-M-$\Gamma$ path (see the left panel). The inset in Fig.~\ref{TBF} zooms in the Y-M-X region (exactly the $(\pi,\pi/6)-(\pi,-\pi/6)$ path). This gives the  bilayer splitting in ($\pi$,0)-point of 160 meV and the absolute bare Fermi velocities: $u_F$(A, bonding) = 2.8 eV\AA, $u_F$(A, antibonding) = 1.0 eV\AA, while $u_F$(N) = 4.0 eV\AA.  

\begin{figure*}
\includegraphics[width=9.6cm]{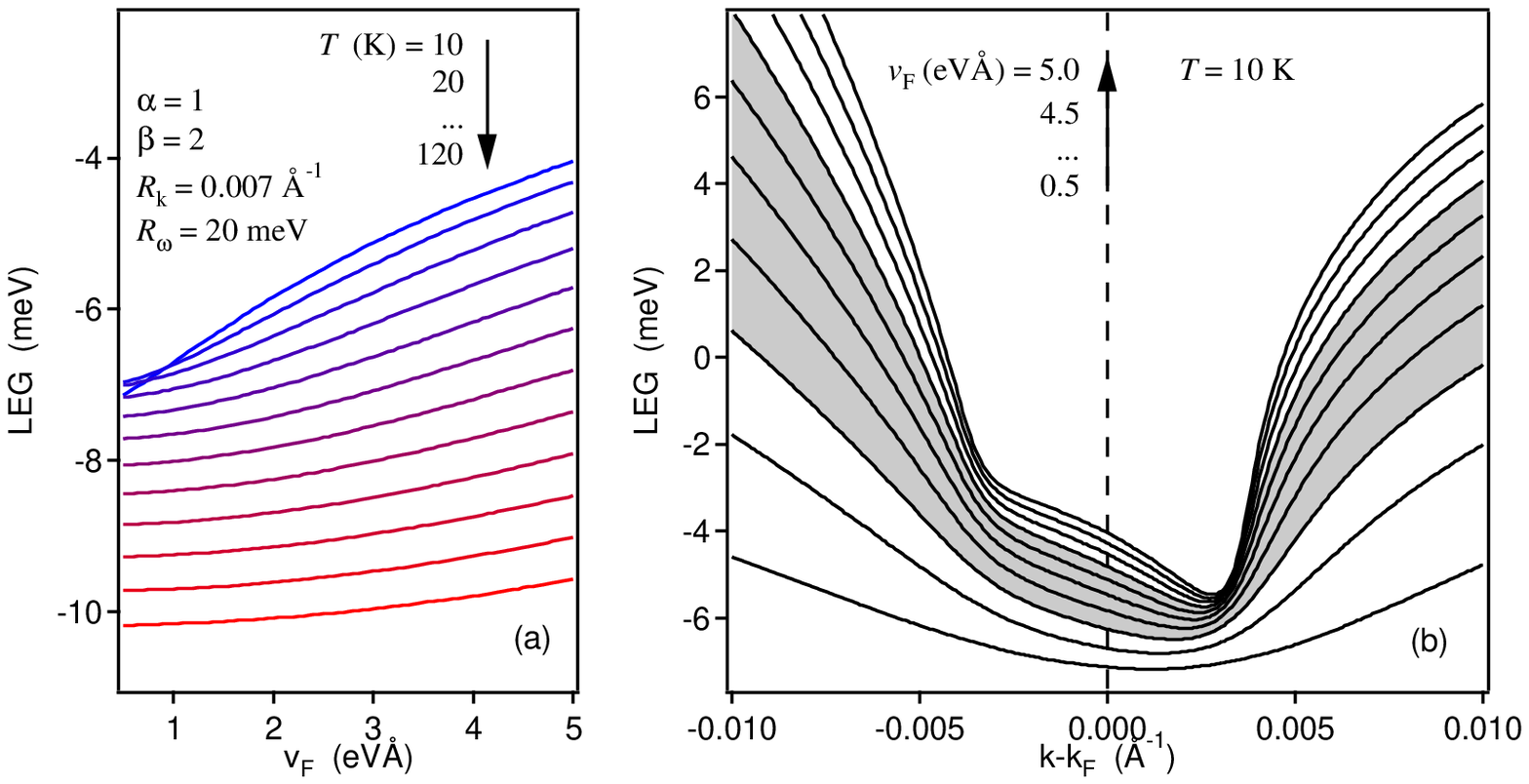}%
\includegraphics[width=8.6cm]{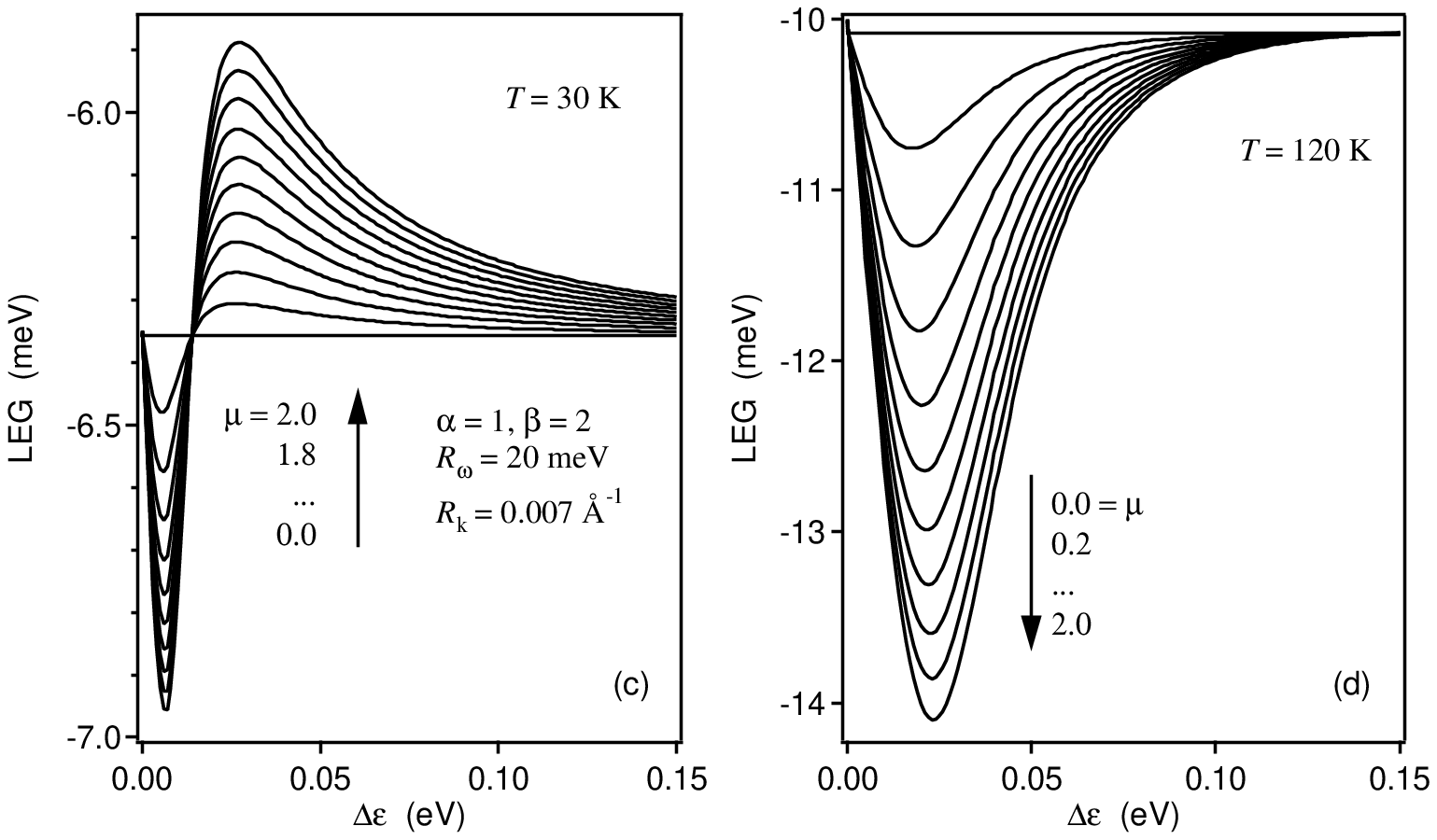}%
\caption{\label{Disp} Dependence of the leading edge position on the band structure parameters: (\textit{a,b}) renormalized Fermi velocity, $v_F = u_F/(1+\lambda)$; and (\textit{c,d}) bilayer splitting, $\Delta \varepsilon = \varepsilon_a - \varepsilon_b$, $\mu = M_a/M_b$ (see Eq.~\ref{splitting}), and $k = k_F$(bonding).}
\end{figure*}

One can expect that the renormalized Fermi velocity $v_F = u_F/(1+\lambda)$ varies even more strongly than the bare one going from the node to the antinode due to a change of the 'coupling strength' $\lambda$.\cite{lambda,GromkoXXX02} So, going back to the simulation procedure, we examine an effect of $v_F$ variation on the leading edge position in the range from 0.5 to 5 eV\AA. The results are represented in Fig.~\ref{Disp} (\textit{a,b}).

\begin{table}
\caption{\label{tab}Tight-binding parameters of the CuO conducting band of Bi-2212.}
\begin{ruledtabular}
\begin{tabular}{lccccc}
Sample & $t$ (eV)& $t'$ (eV)& $t''$ (eV)& $t_{\perp}$ (eV)& $\Delta \epsilon$ (eV)\\
\tableline
OD 69 K & 0.40 & 0.090 & 0.045 & 0.082 & 0.43\\
UD 77 K & 0.39 & 0.078 & 0.039 & 0.082 & 0.29 \\
\end{tabular}
\end{ruledtabular}
\end{table}

From Fig.~\ref{Disp} (\textit{a}) one can see that the influence of $v_F$ on LEG increases with the decreasing temperature but remains rather weak -- about 4 meV change at 10 K over the whole examined range (0.5--5 eV\AA) or in the more reasonable range from 1.5 to 3.5 eV{\AA} (shadowed on panel \textit{b}) inferred from the experiment,\cite{KordPRB02} this variation is only 2 meV for $k = k_F$. Fig.~\ref{Disp} (\textit{b}) also shows that if one follows the minimum gap locus (MGL) criteria~\cite{DingPRL97}---choosing the location in $k$ for which the LEG($k$) along a FS cut has a minimum---this change is even less but, if $k$ deviates more ($|k-k_F| > R_k/2$ = 0.035 \AA$^{-1}$), the change rapidly increases. 

The influence of the bilayer splitting on the leading edge is shown in Fig.~\ref{Disp} (\textit{c,d}). Here $\Delta \varepsilon = \varepsilon_a - \varepsilon_b$ is the value of the splitting, and $\mu = M_a/M_b$ (see Eq.~\ref{splitting}) and $k = k_F$ of the bonding band. One can see that at 120 K the influence of other FS sheet (antibonding in this case) can produce some minor variations (within about 2 meV) in the LEG around the FS but at low temperature (30 K) these variations are negligible.

\subsection{Self-energy}

\begin{figure}
\includegraphics[width=8.2cm]{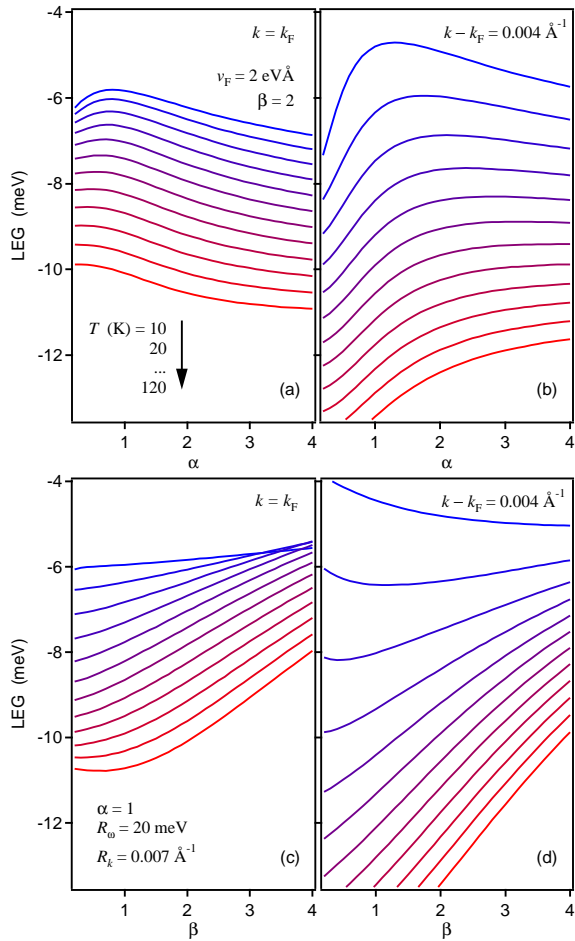}%
\caption{\label{Self} The leading edge gap vs the self-energy parameters $\alpha$ (\textit{a,b}) and $\beta$ (\textit{c,d}). On all panels the different curves correspond to different temperatures as it is shown on panel (\textit{a}).}
\end{figure}

The influence of the self-energy parameters on the leading edge position is presented in Fig.~\ref{Self}. Although, due to the presence of the bilayer splitting, there is no systematic information currently available about how these parameters change over the FS, one can expect that in the antinodal point the coupling strength is larger than in the nodal one and, therefore, the self-energy parameters $\alpha$ and $\beta$ (see above) can be a few times bigger---two times seems to be a reasonable estimation.\cite{BorisPRB01,KordPRL}

Fig.~\ref{Self} (\textit{a}) shows that the variation of $\alpha$ even in a much wider range (from 0.2 to 4) moves the leading edge within about 1 meV only, if $k = k_F$. Small deviation from the $k_F$ noticeably increases the amplitude of the LEG($\alpha$) dependencies (Fig.~\ref{Self} (\textit{b})).

The dependencies of the LEG on $\beta$ within the same range, being more temperature dependent, are similar in amplitude (Fig.~\ref{Self} (\textit{c,d})). 

\section{Gapped case}

\subsection{Minimum gap locus}

\begin{figure}
\includegraphics[width=8.2cm]{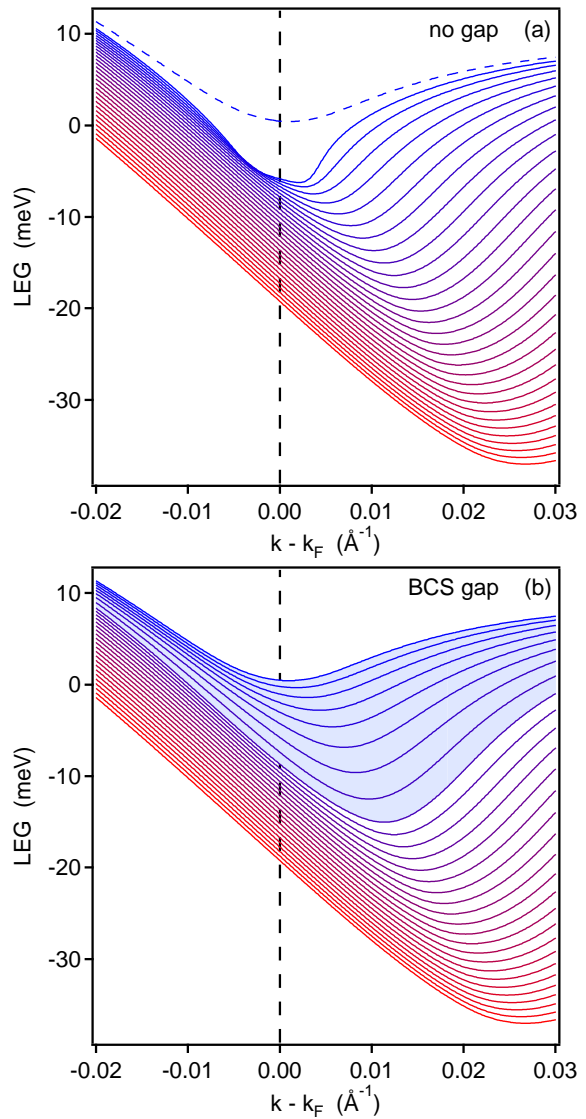}%
\caption{\label{MGL} The dependencies of the leading edge position on $k$ along cut perpendicular to the FS for temperatures from 10 to 300 K (from top to bottom) with the 10 K step for typical for Bi-2212 parameters ($\alpha$ = 1, $\beta$ = 2, $v_F$ = 2~eV\AA, $R_{\omega}$ = 20 meV, $R_k$ = 0.07~\AA$^{-1}$): panel (\textit{a}) represents the gapless case, in panel (\textit{b}) the BCS-like gap opens at $T_c$ = 90 K (the gapped region shadowed with the blue color).}
\end{figure}

\begin{figure*}[tb]
\includegraphics[width=17cm]{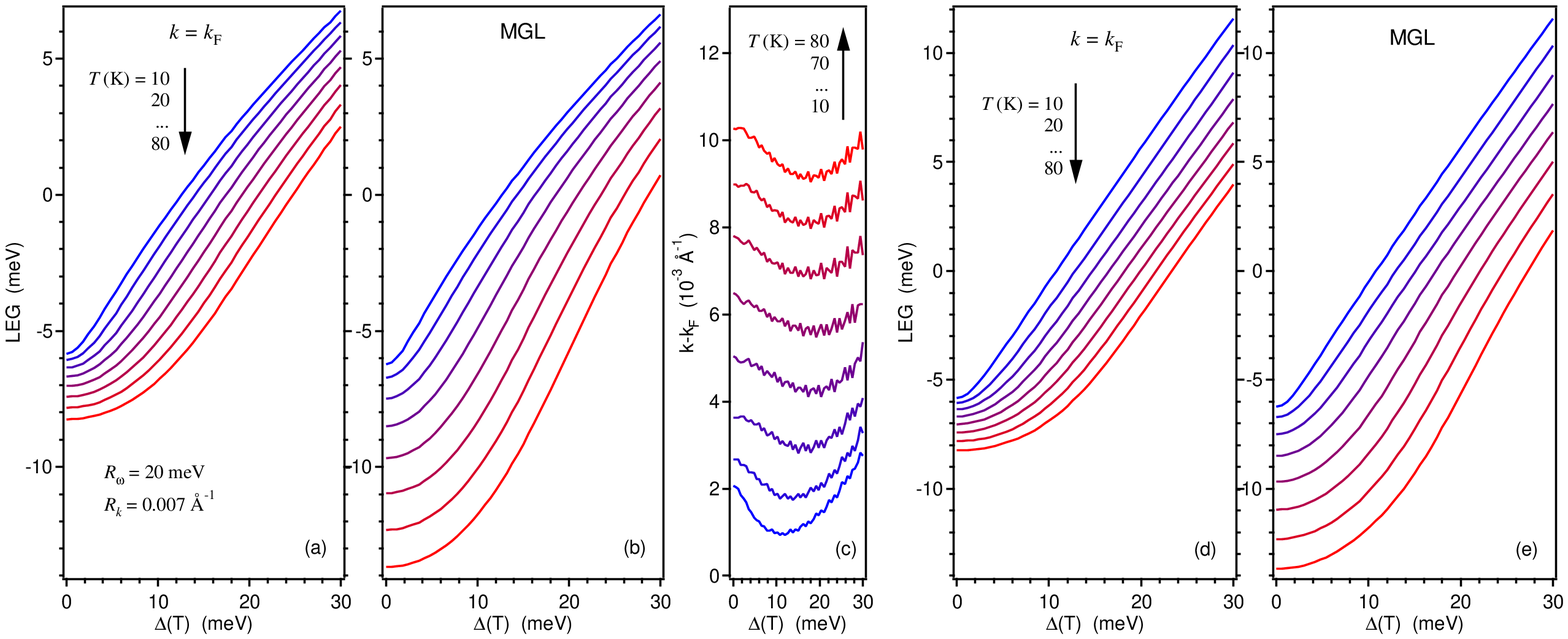}%
\caption{\label{Gap} The leading edge gap (LEG) and the minimum gap locus (MGL) vs real gap value $\Delta(T)$ for the models with the $\Delta$-independent (\textit{a--c}) and BCS-like (\textit{d,e}) self-energies.}
\end{figure*}

Among the $k_F$ determination criteria, the \textquotedblleft minimum gap locus\textquotedblright~(MGL) intuitively seems to be the most suitable in a gapped state. On the other hand, as it is shown above, the MGL, if applied to the $k_F$ determination, gives overestimated values. What is important here is that the $k$-locations which come from the MGL-criterion, $k_{MGL}$, are in the \textquotedblleft dangerous\textquotedblright~region, where a strong dependence of the LEG on temperature and other parameters is expected. We apply the simulation to check at which conditions the MGL method can be used. 

Fig.~\ref{MGL} shows the dependencies of the leading edge position on $k$ along a cut perpendicular to the FS for temperatures from 10 to 300 K (from top to bottom) with 10 K steps. We use parameters typical for the $\Gamma$-X crossing: $\alpha$ = 1, $\beta$ = 2, $v_F$ = 2~eV\AA, $R_{\omega}$ = 20 meV, $R_k$ = 0.07~\AA$^{-1}$. Fig.~\ref{MGL} (\textit{a}) represents the gapless case and Fig.~\ref{MGL} (\textit{b}) shows the same but taking into account a gap which opens at $T_c$ = 90 K. We used the BCS modified spectral function
\begin{eqnarray}\label{GapedA}
A \propto \frac{u_{\textbf{k}}^2 \Sigma''}{(\omega + E_{\textbf{k}})^2 + {\Sigma''}^2} + \frac{(1-u_{\textbf{k}}^2) \Sigma''}{(\omega - E_{\textbf{k}})^2 + {\Sigma''}^2},
\end{eqnarray}
where $E_{\textbf{k}} = \sqrt{\varepsilon_{\textbf{k}}^2 + \Delta(T)^2}$, $u_{\textbf{k}}^2 = \frac{1}{2}(1+\varepsilon_{\textbf{k}}/E_{\textbf{k}})$, and the gap function for $T < T_c$ is approximated as
\begin{eqnarray}\label{MyGap}
\Delta(T) = \Delta_0 \left[1 + \left(1.74 \sqrt{1 - \frac{T}{T_c}}\right)^{2n} \right]^{\frac{1}{2n}}
\end{eqnarray}
with $\Delta_0 = 1.76~T_c$ (temperature in energy units) and $n = -4$.

The gapped region is shadowed in Fig.~\ref{MGL} (\textit{b}) with the blue color, and the gapped LEG($k$, 10 K) dependency is also shown in Fig.~\ref{MGL} (\textit{a}) as a dashed curve to compare with the gapless case.

From the results presented in Fig.~\ref{MGL} one can make the following conclusions. (1) The value of $k$ at which the leading edge position reaches a minimum, $k_{MGL}$, is strongly temperature dependent---thus it is important to realize that the MGL spectra measured at different temperatures correspond to \textit{rather different} $k$-points. (2) The difference between $k_{MGL}$ and $k_F$ is improperly big at high temperatures (above $T_c$) and remains quite big at lower temperatures for both gapped and non-gapped cases. Using a criterion $|k_{MGL}-k_F| < R_k/2$ one can conclude that the MGL method cannot be used for precise $k_F$ determination for $T >$ 40 K, at least in the Bi-2212 case. (3) If one stays at a certain $k$-point, measuring the LEG as a function of temperature, then the region where the real gap opening can be distinguished from the artificial one can be estimated as $k < k_F + R_k/2$. In this case the maximum intensity method,\cite{AebiPRL94} which underestimates the $k_F$,\cite{BorisPRB01} seems to be the most suitable. (4) At low temperatures ($T \leq 40 K$) the MGL method is quite good as a method of the LEG determination---the experimental uncertainty in $k_{MGL}$ is big here because of very flat LEG($k$) dependence but, as another consequence of this, the LEG uncertainty should be small.   

\subsection{LEG in the case of real gaps}

Now we consider how the real gap can influence the LEG value. Fig.~\ref{Gap} represents the results. Panels (\textit{a}) and (\textit{d}) correspond to $k_F$-spectra and panels (\textit{b}) and (\textit{e}) show the LEG of the spectra from $k = k_{MGL}$. The $k_{MGL}(\Delta)$ dependencies at different temperatures are shown in panel (\textit{c}). Note, that all values are shown as function of $\Delta(T)$ rather than $\Delta_0$.

If we accept the use of Eq.~\ref{GapedA} for the gapped spectral function the question how the imaginary part of the self-energy depends on the gap is unclear---in other words, the $\Sigma''(\Delta)$ function is considered to be \textquotedblleft more model dependent\textquotedblright~than the spectral function $A(\Delta)$. The LEG($\Delta$) dependencies presented in Fig.~\ref{Gap} (\textit{a,b}) have been obtained with a $\Delta$-independent self-energy. In Fig.~\ref{Gap} (\textit{d,e}) we model the influence of the gap on the imaginary part of the self energy as $\Sigma''(\omega,\Delta,T) = \sqrt{(\alpha_g \omega)^2 + (\beta T)^2}$, where  
\begin{eqnarray}\label{S3}
\alpha_g(\omega,\Delta) = \alpha
	\begin{cases}
	\omega^2\Delta^{-2} & \text{if $|\omega| < \Delta$}, \\
	1 & \text{if $|\omega| > \Delta$},
	\end{cases}
\end{eqnarray}
which gives the BCS asymptotic $\Sigma''(\omega) \propto \omega^3$ at $\omega \rightarrow 0$ and zero temperature.

One can see that there is some \textquotedblleft rounding\textquotedblright~(deviation from linear) of the LEG($\Delta$) curves at low $\Delta$ which is explained by cutting finite width EDCs by the Fermi-function. These \textquotedblleft rounding\textquotedblright~increases with temperature but it is possible to conclude that for low temperatures ($T <$ 30 K) it remains rather small (less than 2 meV) for all $k$-points and $\Sigma''(\omega)$ models considered. This means, for example, that the U-shape of the LEG($\varphi$)~\cite{phi} observed in Ref.~\onlinecite{BorisXXX02} cannot be explained by such a rounding of a sharp cusp expected in a simplest $d$-wave gap function.

The comparison of Fig.~\ref{Gap} (\textit{a,b}) and (\textit{d,e}) respectively shows that at higher $\Delta$ the LEG($\Delta$) dependencies become model dependent. Then, from the \textquotedblleft more physical\textquotedblright~model of $\Sigma''(\omega)$, determined by Eq.~\ref{S3}, we can get a rough estimation of the coefficient between the LEG and $\Delta$: $d$LEG($\Delta$)$/d\Delta \approx 0.5$.  

\section{Discussion}

With the above simulations we mostly focused on the gapless case---we examined the leading edge position of the photoemission spectra in case when there is no gap present in the electron density of states. 

It has appeared that the leading edge position of such a gapless spectrum depends on temperature, resolutions, band structure and self-energy parameters, which is not surprising of course. A surprising result is that under certain circumstances these dependencies can  be quite complicated (exhibit non-monotonic rate) that can be misinterpreted as a physical transition. 

These \textquotedblleft circumstances\textquotedblright~are a wrong position in momentum space, i.e.~when $k$-location of the spectrum does not coincide with $k_F$ but is uncertain. If the parameters mentioned above are well determined experimentally and can be taken into account with the described simulation, the uncertainty in momentum cannot be dealt with in the same way, by definition. More precisely, this uncertainty is the uncertainty in $k - k_F$ and, therefore, consists of the two: $k$-uncertainty, which is determined by the $k$ step of the measurements, and $k_F$-uncertainty which is related to a $k_F$ determination procedure. The first one can be reduced by refining the $k$ step of the measurements, which should be significantly less that the momentum resolution. The second problem is well known and heavily discussed in the past but we would like to address this question a bit more. 

One way to define an uncertainty of the method is to model how far the directly determined value of $k_F$ deviates from the real one---for example, in the \textquotedblleft maximum intensity\textquotedblright~method, how the value of $k$, at which the photocurrent intensity exhibits a maximum, deviates from the real $k_F$. It has been shown that for high temperatures the maximum intensity method, especially in combination with the normalization to \textquotedblleft highest binding energy", is much more precise than others, such as \textquotedblleft $\nabla n(\textbf{k})$\textquotedblright, \textquotedblleft minimum gap locus\textquotedblright~or \textquotedblleft $\Delta T$\textquotedblright~methods,\cite{BorisPRB01,KordPRB02} but modelling this, one can find that it is difficult to suggest such a universal procedure of the $k_F$ determination which always gives the best result. All of the methods give some deviations and then one can reason that it is possible to determine the real $k_F$ taking these deviations into account turning the uncertainty to zero. Then we come to another uncertainty: how much these deviations depend on the model of the spectral function we use. This is an interesting and complicated question which can be subject of a separate study. Here we apply the word \textquotedblleft uncertainty\textquotedblright~of the method in the first sense and, in this respect, the maximum intensity method appeares to be the most accurate.

What is the real benefit one can extract from the results obtained or what do these results contribute to the gap problem? There are two kinds of gap measurements which are considered to be \textquotedblleft model independent\textquotedblright: the gap as a function of temperature, LEG($T$), and momentum, LEG($\textbf{k}$), which is usually presented as a function of FS angle, LEG($\varphi$). In both cases the obtained \textquotedblleft gap\textquotedblright~is relative: in the first case it is relative to the LEG at the highest temperature, in the second case it is relative to the LEG in nodal direction. 

The presented simulation shows that even when a real gap is zero the LEG is strongy temperature dependent (possibly in a complex way), and the amplitude of this dependency is comparable with the estimated real gap values in the superconducting cuprates. On the other hand, the momentum dependence of such a gapless LEG at low temperatures is shown to be rather weak for all possible variations of physical, structural and experimental parameters. But what is most important is that knowing these parameters one can estimate the absolute leading edge position and, therefore, answer the question whether there is any gap in a given spectrum or not. 

For example, the LEG at 40 K and with the other parameters used in Ref.~\onlinecite{BorisXXX02} for the nodal direction is equal to $-7$ meV (see Fig.~\ref{Tdep}, \textit{d}--\textit{f}) which precisely coincides with the value obtained in the cited paper. This fact allowed us to make a firm conclusion that the nodal direction in an underdoped Bi-2212 in superconducting state is really the node---the gap is \textit{zero}---which is in favor of $d$-symmetry of both superconducting and pseudo-gaps.

Another conclusion comes from Fig.~\ref{Gap}: the \textquotedblleft U-shape\textquotedblright~of the LEG($\varphi$) dependencies, which have been recenly observed for underdoped samples,~\cite{BorisXXX02} cannot be explained neither by the interplay between the temperature and Fermi-function nor by other \textquotedblleft artificial\textquotedblright~parameters which may vary over the FS: momentum resolution, Fermi velocity, self-energy parameters. Considering the absence of the gap in the node as an evidence for $d$-wave symmetry of the order parameter in Bi-2212, this U-shape can be a consequence of higher harmonics of the gap due to the long range of the pairing interaction, as it was suggested earlier,\cite{MesotPRL99} or due to an interplay between the superconducting and pseudo-gaps. We want to note that this conclusion is valid for low temperatures only (superconducting state for Bi-2212), for higher temperatures (e.g. for pseudo-gap state) the influence of the artificial parameters can be essential.  

\section{Summary}

In conclusion, a systematic study of the LEG is presented. Even in the absence of a real gap, the leading edge position of the ARPES spectrum depends on many factors: energy and momentum resolutions, Fermi-velocity, the self-energy, but mostly on temperature. At some circumstances, when the spectra are measured away from $k_F$ (if $k - k_F >$ 0.005 \AA$^{-1}$ for typical parameters of an experiment on Bi-2212), the temperature dependence of the LEG can be quite complicated and exhibit non-monotonic rate that can be misinterpreted as a physical transition. At low temperatures the dependence of the LEG on other parameters is rather weak and cannot be responsible for observed in experiments the LEG variation over the FS. 

The \textit{absolute} gap values which have been derived from the presented simulation prove that the nodal direction in the underdoped Bi-2212 in the superconducting state is really the node (the gap is \textit{zero}) but also that the \textquotedblleft U-shape\textquotedblright~of the LEG vs Fermi surface angle, observed for underdoped samples, is not artificial. This implies the $d$-wave symmetry of the superconducting and pseudo-gaps but with essential contributions from the higher harmonics of one or both gaps which can be a result of the long range pairing interaction.

\begin{acknowledgments}
We acknowledge the stimulating discussions with O.~K.~Andersen, M.~Eschrig, E.~Laptev, and C.~Varma.
\end{acknowledgments}


\begin{thebibliography}{}
\bibitem{Self-energy} We note that any similarity of this empirical formula to existing theoretical models should be considered as a coincidence.
\bibitem{lambda} Note, that $\lambda$ here includes all interactions which result in the finite self-energy of the quasiparticles.
\end{thebibliography}
\end{document}